\documentclass[journal,english]{IEEEtran}
\usepackage{float}
\usepackage{amsmath}
\usepackage{graphicx}

\makeatletter

\providecommand{\tabularnewline}{\\}
\floatstyle{ruled}
\newfloat{algorithm}{tbp}{loa}
\floatname{algorithm}{Algorithm}

\usepackage{cite}
\usepackage{url}

\usepackage{algorithmic}

\newcommand{\fftfunc}[3]{\operatorname{\mathbf{#1}}_{#2}\left(#3\right)}

\newcommand{\fftfuncl}[4]
{\operatorname{\mathbf{#1}}_{#2}^{#3}\left(#4\right)}

\newcommand{\nsplitfftdsfour}[2]
{\fftfuncl{newfftS}{#1}{4}{#2}}
\newcommand{\nsplitfftdsl}[3]
{\fftfuncl{newfftS}{#1}{#2}{#3}}

\newcommand{\nsplitdctIII}[3]
{\fftfuncl{newdctIII}{#1}{#2}{#3}}
\newcommand{\nsplitdstIII}[3]
{\fftfuncl{newdstIII}{#1}{#2}{#3}}

\newcommand{\nsplitdctIV}[2]
{\fftfunc{newdctIV}{#1}{#2}}

\newcommand{\dft}{\operatorname{\mathrm{dft}}}

\renewcommand{\Re}{\operatorname{Re}}
\renewcommand{\Im}{\operatorname{Im}}

\usepackage{babel}
\makeatother
\begin{document}

\title{Type-IV DCT, DST, and MDCT algorithms\\
with reduced numbers of arithmetic operations}

\author{Xuancheng Shao and Steven G. Johnson{*}% <-this % stops a space
\thanks{* Department of Mathematics, Massachusetts Institute of Technology,
Cambridge MA 01239.}%
}

\maketitle

\begin{abstract}
We present algorithms for the type-IV discrete cosine transform
(DCT-IV) and discrete sine transform (DST-IV), as well as for the
modified discrete cosine transform (MDCT) and its inverse, that
achieve a lower count of real multiplications and additions than
previously published algorithms, without sacrificing numerical
accuracy. Asymptotically, the operation count is reduced from
$2N\log_{2}N+O(N)$ to $\frac{17}{9}N\log_{2}N+O(N)$ for a power-of-two
transform size $N$, and the exact count is strictly lowered for all
$N\ge8$. These results are derived by considering the DCT to be a
special case of a DFT of length $8N$, with certain symmetries, and
then pruning redundant operations from a recent improved fast Fourier
transform algorithm (based on a recursive rescaling of the
conjugate-pair split-radix algorithm). The improved algorithms for
DST-IV and MDCT follow immediately from the improved count for the
DCT-IV.
\end{abstract}

\section{Introduction}

In this paper, we present recursive algorithms for type-IV discrete
cosine and sine transforms (DCT-IV and DST-IV) and modified discrete
cosine transforms (MDCTs), of power-of-two sizes $N$, that require
fewer total real additions and multiplications (herein called \emph{flops})
than previously published algorithms (with an asymptotic reduction
of about 6\%), without sacrificing numerical accuracy. This work,
extending our previous results for small fixed $N$ \cite{Johnson07},
appears to be the first time in over 20 years that flop counts for
the DCT-IV and MDCT have been reduced---although computation times
are no longer generally determined by arithmetic counts \cite{FFTW05},
the question of the minimum number of flops remains of fundamental
theoretical interest. Our fast algorithms are based on one of two recently
published fast Fourier transform (FFT) algorithms~\cite{Johnson07,Lundy07}, which reduced the
operation count for the discrete Fourier transform (DFT) of size $N$
to $\frac{34}{9}N\log_{2}N+O(N)$ compared to the
(previous best) split-radix algorithm's $4N\log_{2}N+O(N)$ \cite{Yavne68,Duhamel84,Martens84,Vetterli84,DuhamelVe90}.
Given the new FFT, we treat a DCT as an FFT of real-symmetric inputs
and eliminate redundant operations to derive the new algorithm; in
other work, we applied the same approach to derive improved algorithms
for the type-II and type-III DCT and DST \cite{ShaoJo07-preprint}. 

The algorithm for DCT-IV that we present has the same recursive structure
as some previous DCT-IV algorithms, but the subtransforms are recursively
rescaled in order to eliminate some of the multiplications. This approach
reduces the flop count for the DCT-IV from the previous best of $2N\log_{2}N+N$
\cite{Wang85,Suehiro86,Chan90,Kok97,Britanak99,Plonka02,Britanak03}
to:\begin{equation}
\frac{17}{9}N\log_{2}N+\frac{31}{27}N+\frac{2}{9}(-1)^{\log_{2}N}\log_{2}N-\frac{4}{27}(-1)^{\log_{2}N}.\label{eq:flop-count-dct-IV}\end{equation}
The first savings occur for $N=8$, and are summarized in Table\@.~\ref{tab:counts}.%
\begin{table}
\begin{centering}\begin{tabular}{|c|c|c|}
\hline 
$N$&
previous DCT-IV&
New algorithm\tabularnewline
\hline
\hline 
8&
56&
54\tabularnewline
\hline 
16&
144&
140\tabularnewline
\hline 
32&
352&
338\tabularnewline
\hline 
64&
832&
800\tabularnewline
\hline 
128&
1920&
1838\tabularnewline
\hline 
256&
4352&
4164\tabularnewline
\hline 
512&
9728&
9290\tabularnewline
\hline 
1024&
21504&
20520\tabularnewline
\hline 
2048&
47104&
44902\tabularnewline
\hline 
4096&
102400&
97548\tabularnewline
\hline
\end{tabular}\par\end{centering}

\caption{\label{tab:counts}Flop counts (real adds + mults) of previous best
DCT-IV and our new algorithm }
\end{table}
 In order to derive a DCT-IV algorithm from the new FFT algorithm,
we simply consider the DCT-IV to be a special case of a DFT with real
input of a certain symmetry, and discard the redundant operations.
{[}This should not be confused with algorithms that employ an \emph{unmodified}
FFT combined with $O(N)$ pre/post-processing steps to obtain the
DCT.] This well-known technique \cite{Vetterli84,Duhamel86,DuhamelVe90,VuducDe00,FFTW05,Johnson07,ShaoJo07-preprint}
allows any improvements in the DFT to be immediately translated to
the DCT-IV, is simple to derive, avoids cumbersome re-invention of
the {}``same'' algorithm for each new trigonometric transform, and
(starting with a split-radix FFT) matches the best previous flop counts
for every type of DCT and DST. The connection to a DFT of symmetric
data can also be viewed as the basic reason why DCT flop counts had
not improved for so long: as we review below, the old DCT flop counts
can be derived from a split-radix algorithm \cite{Plonka02}, and
the 1968 flop count of split radix was only recently improved upon
\cite{Johnson07,Lundy07}. There have been many previously published
DCT-IV algorithms derived by a variety of techniques, some achieving
$2N\log_{2}N+N$ flops \cite{Wang85,Suehiro86,Chan90,Kok97,Britanak99,Plonka02,Britanak03}
and others obtaining larger or unreported flop counts \cite{Wang84,Murthy90,Puschel03,Plonka05}.
Furthermore, exactly the same flop count (\ref{eq:flop-count-dct-IV})
is now obtained for the type-IV discrete sine transform (DST-IV),
since a DST-IV can be obtained from a DCT-IV via flipping the sign
of every other input (zero flops, since the sign changes can be absorbed
by converting subsequent additions into subtractions or vice versa)
and a permutation of the output (zero flops) \cite{Chan90}. Also,
in many practical circumstances the output can be scaled by an arbitrary
factor (since any scaling can be absorbed into a subsequent computation);
in this case, similar to the well-known savings for a scaled-output
size-8 DCT-II in JPEG compression \cite{Arai88,Pennebaker93,Johnson07,ShaoJo07-preprint},
we show that an additional $N$ multiplications can be saved for a
scaled-output (or scaled-input) DCT-IV.

Indeed, if we only wished to show that the asymptotic flop count for
DCT-IV could be reduced to $\frac{17}{9} N \log_2 N+O(N)$, we could simply
apply known algorithms to express a DCT-IV in terms of a real-input
DFT (e.g. by reducing it to the DCT-III~\cite{Wang85} and thence to a
real-input DFT~\cite{Narasimha78}) to immediately apply the $\frac{17}{9}
N\log_2 N+O(N)$ flop count for a real-input DFT from our previous
paper~\cite{Johnson07}.  However, with FFT and DCT algorithms, there
is great interest in obtaining not only the best possible asymptotic
constant factor, but also the best possible \emph{exact} count of
arithmetic operations. Our result (\ref{eq:flop-count-dct-IV}) is
intended as a new upper bound on this (still unknown) minimum exact
count, and therefore we have done our best with the $O(N)$ terms as
well as the asymptotic constant.

An important transform closely related to the DCT-IV is an MDCT, which
takes $2N$ inputs and produces $N$ outputs, and is designed to be
applied to 50\%-overlapped blocks of data \cite{Princen86,Johnson87}.
Such a {}``lapped'' transform reduces artifacts from block boundaries
and is widely used in audio compression \cite{Painter00}. In fact,
an MDCT is \emph{exactly} equivalent to a DCT-IV of size $N$, where
the $2N$ inputs have been preprocessed with $N$ additions/subtractions
\cite{Malvar90,Malvar91,Chan96,Liu99}. This means that the flop count
for an MDCT is at most that of a DCT-IV plus $N$ flops. Precisely
this technique led to the best previous flop count for an MDCT, $2N\log_{2}N+2N$
\cite{Malvar90,Malvar91,Chan96,Liu99}. (There have also been several
MDCT algorithms published with larger or unreported flop counts \cite{Malvar92,Sporer92,Chiang96,Fan99,Lee01,Jing01,Britanak02,Cheng03,Nikolajevic03,Chen03}.)
It also means that our improved DCT-IV immediately produces an improved
MDCT, with a flop count of eq\@.~(\ref{eq:flop-count-dct-IV}) plus
$N$. Similarly for the inverse MDCT (IMDCT), which takes $N$ inputs
to $2N$ outputs and is equivalent to a DCT-IV of size $N$ plus $N$
negations (which should not, we argue, be counted in the flops because
they can be absorbed by converting subsequent additions into subtractions).

In the following sections, we first briefly review the new FFT algorithm,
previously described in detail \cite{Johnson07}. Then, we review
how a DCT-IV may be expressed as a special case of a real DFT, and
how the new DCT-IV algorithm may be derived by applying the new FFT
algorithm and pruning the redundant operations. In doing so, we find
it necessary to develop a algorithm for a DCT-III with scaled output.
(Previously, we had derived a fast algorithm for a DCT-III based on
the new FFT, but only for the case of scaled or unscaled \emph{input}
\cite{ShaoJo07-preprint}.) This DCT-III algorithm follows the same
approach of eliminating redundant operations from our new scaled-output
FFT (a subtransform of the new FFT) applied to appropriate real-symmetric
inputs. We then analyze the flop counts for the DCT-III and DCT-IV
algorithms. Finally, we show that this improved DCT-IV immediately
leads to improved DST-IV, MDCT, and IMDCT algorithms. We close with
some concluding remarks about future directions.

\section{Review of the new FFT }

To obtain the new FFT, we used as our starting point a variation called
the \emph{conjugate-pair FFT} of the well-known split-radix algorithm.
Here, we first review the conjugate-pair FFT, and then briefly summarize
how this was modified to reduce the number of flops.

\subsection{Conjugate-pair FFT\label{sec:Conjugate-pair-FFT}}

The discrete Fourier transform of size $N$ is defined by \begin{equation}
X_{k}=\sum_{n=0}^{N-1}x_{n}\omega_{N}^{nk},\label{eq:DFT}\end{equation}
where $\omega_{N}=e^{-\frac{2\pi i}{N}}$ is an $N$th primitive root
of unity and $k=0,\ldots,N-1$. 

Starting with this equation, the decimation-in-time conjugate-pair
FFT \cite{Kamar89,Johnson07}, a variation on the well-known split-radix
algorithm \cite{Yavne68,Duhamel84,Martens84,Vetterli84}, splits it
into three smaller DFTs: one of size $N/2$ of the even-indexed inputs,
and two of size $N/4$:\begin{multline}
X_{k}=\sum_{n_{2}=0}^{N/2-1}\omega_{N/2}^{n_{2}k}x_{2n_{2}}+\omega_{N}^{k}\sum_{n_{4}=0}^{N/4-1}\omega_{N/4}^{n_{4}k}x_{4n_{4}+1}\\+\,\omega_{N}^{-k}\sum_{n_{4}=0}^{N/4-1}\omega_{N/4}^{n_{4}k}x_{4n_{4}-1}.\label{eq:conjugate-pair}\end{multline}
 {[}In contrast, the ordinary split-radix FFT uses $x_{4n_{4}+3}$
for the third sum (a cyclic shift of $x_{4n_{4}-1}$), with a corresponding
multiplicative {}``twiddle'' factor of $\omega_{N}^{3k}$.] This
decomposition is repeated recursively until base cases of size $N=1$
or $N=2$ are reached. The number of flops required by this algorithm,
after certain simplifications (common subexpression elimination and
constant folding) and not counting data-independent operations like
the computation of $\omega_{N}^{k}$, is $4N\log_{2}N-6N+8$, identical
to ordinary split radix \cite{Gopinath89,Qian90,Krot92,Johnson07}.

\subsection{New FFT\label{sub:newfft}}

Based on the conjugate-pair split-radix FFT from section~\ref{sec:Conjugate-pair-FFT},
a new FFT algorithm with a reduced number of flops can be derived
by scaling the subtransforms \cite{Johnson07}. We will not reproduce
the derivation here, but will simply summarize the results. In particular,
the original conjugate-pair split-radix algorithm is split into four
mutually recursive algorithms, $\ \nsplitfftdsl N\ell x$ for $\ell=0,1,2,4$,
each of which has the same split-radix structure but computes a DFT
scaled by a factor of $1/s_{\ell N,k}$ (defined below), respectively.
These algorithms are shown in Algorithm~\ref{alg:new-fft}, in which
the scaling factors are combined with the twiddle factors $\omega_{N}^{k}$
to reduce the total number of multiplications. In particular, all
of the savings occur in $\ \nsplitfftdsl N1x$, while $\ \nsplitfftdsl N4x$
is factorized into a special form to minimize the number of extra
multiplications it requires. Here, although $\ell=0,1,2$ are presented
in a compact form by a single subroutine $\ \nsplitfftdsl N\ell x$,
in practice they would have to be implemented as three separate subroutines
in order to exploit the special cases of the multiplicative constants
$s_{N,k}/s_{\ell N,k}$, as described in Ref\@.~\cite{Johnson07}.
For simplicity, we have omitted the base cases of the recursion ($N=1$
and $2$) and have not eliminated common subexpressions. %
\begin{algorithm}
\begin{algorithmic}

\item[\textbf{function}] $X_{k=0..N-1} \leftarrow \nsplitfftdsl{N}{\ell}{x_n}$:

\COMMENT{computes DFT / $s_{\ell N,k}$, $\ell=0,1,2$}

\STATE $U_{k_2=0 \ldots N/2-1} \leftarrow \nsplitfftdsl{N/2}{2\ell}{x_{2n_2}}$

\STATE $Z_{k_4=0 \ldots N/4-1} \leftarrow \nsplitfftdsl{N/4}{1}{x_{4n_4+1}}$

\STATE $Z'_{k_4=0 \ldots N/4-1} \leftarrow \nsplitfftdsl{N/4}{1}{x_{4n_4-1}}$

\FOR{$k=0$ to $N/4-1$}

\STATE $X_k \leftarrow U_k + \left( t_{N,k} Z_k + t_{N,k}^* Z'_k \right) \cdot (s_{N,k} / s_{\ell N,k})$

\STATE $X_{k+N/2} \leftarrow U_k - \left(t_{N,k} Z_k + t_{N,k}^* Z'_k \right)\cdot (s_{N,k} / s_{\ell N,k})$

\STATE $X_{k+N/4} \leftarrow U_{k+N/4} $ \\ \hspace{1em} $ - \: i \left(t_{N,k} Z_k - t_{N,k}^* Z'_k \right)\cdot (s_{N,k} / s_{\ell N,k+N/4})$

\STATE $X_{k+3N/4} \leftarrow U_{k+N/4} $ \\ \hspace{1em} $ + \: i \left(t_{N,k} Z_k - t_{N,k}^* Z'_k \right)\cdot (s_{N,k} / s_{\ell N,k+N/4})$

\ENDFOR

\vskip 3pt

\item[\textbf{function}] $X_{k=0..N-1} \leftarrow \nsplitfftdsfour{N}{x_n}$:

\COMMENT{computes DFT / $s_{4N,k}$}

\STATE $U_{k_2=0 \ldots N/2-1} \leftarrow \nsplitfftdsl{N/2}{2}{x_{2n_2}}$

\STATE $Z_{k_4=0 \ldots N/4-1} \leftarrow \nsplitfftdsl{N/4}{1}{x_{4n_4+1}}$

\STATE $Z'_{k_4=0 \ldots N/4-1} \leftarrow \nsplitfftdsl{N/4}{1}{x_{4n_4-1}}$

\FOR{$k=0$ to $N/4-1$}

\STATE $X_k \leftarrow \left[ U_k + \left( t_{N,k} Z_k + t_{N,k}^* Z'_k \right) \right] \cdot (s_{N,k} / s_{4N,k})$

\STATE $X_{k+N/2} \leftarrow \left[ U_k - \left(t_{N,k} Z_k + t_{N,k}^* Z'_k \right) \right] $ \\ \hspace{5em} $ \cdot \: (s_{N,k} / s_{4N,k+N/2})$

\STATE $X_{k+N/4} \leftarrow \left[ U_{k+N/4} - i \left(t_{N,k} Z_k - t_{N,k}^* Z'_k \right) \right] $ \\ \hspace{5em} $ \cdot \: (s_{N,k} / s_{4N,k+N/4})$

\STATE $X_{k+3N/4} \leftarrow \left[ U_{k+N/4} + i \left(t_{N,k} Z_k - t_{N,k}^* Z'_k \right) \right] $ \\ \hspace{5em} $ \cdot \: (s_{N,k} / s_{4N,k+3N/4})$

\ENDFOR

\end{algorithmic}

\caption{\label{alg:new-fft}New FFT algorithm of length $N$ (divisible by
$4$). The sub-transforms $\nsplitfftdsl N\ell x$ for $\ell\neq0$
are scaled by $s_{\ell N,k}$, respectively, while $\ell=0$ is the
final unscaled DFT ($s_{0,k}=1$).}
\end{algorithm}

The key aspect of these algorithms is the scale factor $s_{N,k}$,
where the subtransforms compute the DFT scaled by $1/s_{\ell N,k}$
for $\ell=1,2,4$. This scale factor is defined for $N=2^{m}$ by
the following recurrence, where $k_{4}=k$ mod $\frac{N}{4}$:

\begin{equation}
s_{N=2^{m},k}=\left\{ \begin{array}{c}
1\quad\textrm{for }N\leq4\\
s_{N/4,k_{4}}\cos(2\pi k_{4}/N)\quad\textrm{for }k_{4}\leq N/8\\
s_{N/4,k_{4}}\sin(2\pi k_{4}/N)\quad\textrm{otherwise}\end{array}\right..\label{eq:scalesym}\end{equation}
This definition has the properties: $s_{N,0}=1$, $s_{N,k+N/4}=s_{N,k}$,
and $s_{N,N/4-k}=s_{N,k}$. Also, $s_{N,k}>0$ and decays rather slowly
with $N$: $s_{N,k}$ is $\Omega(N^{\log_{4}\cos(\pi/5)})$ asymptotically
\cite{Johnson07}. When these scale factors are combined
with the twiddle factors $\omega_{N}^{k}$, we obtain terms of the
form \begin{equation}
t_{N,k}=\omega_{N}^{k}\frac{s_{N/4,k}}{s_{N,k}},\label{eq:t}\end{equation}
which is always a complex number of the form $\pm1\pm i\tan\frac{2\pi k}{N}$
or $\pm\cot\frac{2\pi k}{N}\pm i$ and can therefore be multiplied
with two fewer real multiplications than are required to multiply
by $\omega_{N}^{k}$. We denote the complex conjuate of $t_{N,k}$ by $t_{N,k}^*$. The resulting flop count, for arbitrary complex
data $x_{n}$, is then reduced from $4N\log_{2}N-6N+8$ for split
radix to $\frac{34}{9}N\log_{2}N+O(N)$ \cite{Johnson07}.

\section{Fast DCT-IV from new FFT}

Various forms of discrete cosine transform have been defined, corresponding
to different boundary conditions on the transform. The type-IV DCT
is defined as a real, linear transformation by the formula:

\begin{equation}
C_{k}^{\mathrm{IV}}=\sum_{n=0}^{N-1}x_{n}\cos\left[\frac{\pi}{N}\left(n+\frac{1}{2}\right)\left(k+\frac{1}{2}\right)\right],\label{eq:DCT-IV}\end{equation}
 for $N$ inputs $x_{n}$ and $N$ outputs $C_{k}^{\mathrm{IV}}$.
The transform can be made orthogonal (unitary) by multiplying with
a normalization factor $\sqrt{2/N}$, but for our purposes the unnormalized
form is more convenient (and has no effect on the number of operations).
We will now derive an algorithm, starting from the new FFT of the
previous section, to compute the DCT-IV in terms of a scaled DCT-III
and DST-III. These type-III transforms are then treated in following
section by a similar method, and lead to our new flop count for the
DCT-IV.

In particular, we wish to emphasize in this paper that the DCT-IV
(and, indeed, all types of DCT) can be viewed as special cases of the
discrete Fourier transform (DFT) with real inputs of a certain
symmetry. This viewpoint is fruitful because it means that any FFT
algorithm for the DFT leads immediately to a corresponding fast
algorithm for the DCT-IV simply by discarding the redundant
operations, rather than rederiving a {}``new'' algorithm from
scratch. A similar viewpoint has been used to derive fast algorithms
for the DCT-II
\cite{Vetterli84,Duhamel86,DuhamelVe90,VuducDe00,FFTW05,Johnson07}, as
well as in automatic code-generation for the DCT-IV
\cite{FFTW05,Johnson07}, and has been observed to lead to the minimum
known flop count starting from the best known DFT
algorithm. Furthermore, because the algorithm is equivalent to an FFT
algorithm with certain inputs, it should have the same floating-point
error characteristics as that FFT---in this case, the underlying FFT
algorithm is simply a rescaling of split radix~\cite{Johnson07}, and
therefore inherits the favorable $O(\sqrt{\log N})$ mean error growth
and $O(\log N)$ error bounds of the Cooley-Tukey algorithm
\cite{GenSan66,Schatzman96,Tasche00}, unlike at least one other DCT-IV
algorithm \cite{Chan90} that has been observed to display
$O(\sqrt{N})$ error growth \cite{FFTW05}.

\subsection{DCT-IV in terms of DFT\label{sub:DCT-IV-from-DFT}}

Recall that the discrete Fourier transform of size $N$ is defined
by eq.~(\ref{eq:DFT}). In order to derive $C_{k}^{\mathrm{IV}}$
from the DFT formula, one can use the identity $\cos\frac{\pi\ell}{N}=\frac{1}{4}\left(\omega_{8N}^{4\ell}-\omega_{8N}^{4N-4\ell}-\omega_{8N}^{4N+4\ell}+\omega_{8N}^{8N-4\ell}\right)$
to write:\begin{align}
C_{k}^{\mathrm{IV}} & =\sum_{n=0}^{N-1}x_{n}\cos\left[\frac{\pi}{N}\left(n+\frac{1}{2}\right)\left(k+\frac{1}{2}\right)\right]\nonumber \\
 & =\sum_{n=0}^{N-1}\frac{x_{n}}{4}\left[\omega_{8N}^{(2n+1)(2k+1)}-\omega_{8N}^{(4N-2n-1)(2k+1)}\right.\nonumber \\
 & \quad\quad\left.-\,\omega_{8N}^{(4N+2n+1)(2k+1)}+\omega_{8N}^{(8N-2n-1)(2k+1)}\right]\nonumber \\
 & =\sum_{n=0}^{8N-1}\tilde{x}_{n}\omega_{8N}^{n(2k+1)},\label{eq:DCT-IV from DFT}\end{align}
where $\tilde{x}_{n}$ is a real-even sequence of length $\tilde{N}=8N$,
defined as follows for $0\leq n<N$:

\begin{equation}
\tilde{x}_{2n+1}=\tilde{x}_{8N-2n-1}=\frac{1}{4}x_{n}\label{eq:tilde1}\end{equation}

\begin{equation}
\tilde{x}_{4N-2n-1}=\tilde{x}_{4N+2n+1}=-\frac{1}{4}x_{n}\label{eq:tilde2}\end{equation}
Furthermore, the even-indexed inputs are zeros: $\tilde{x}_{2n}=0$
for all $0\leq n<4N$. (The factors of $1/4$ will disappear in the
end because they cancel equivalent factors in the subtransforms.)

\begin{figure}
\begin{centering}\includegraphics[width=1.0\columnwidth]{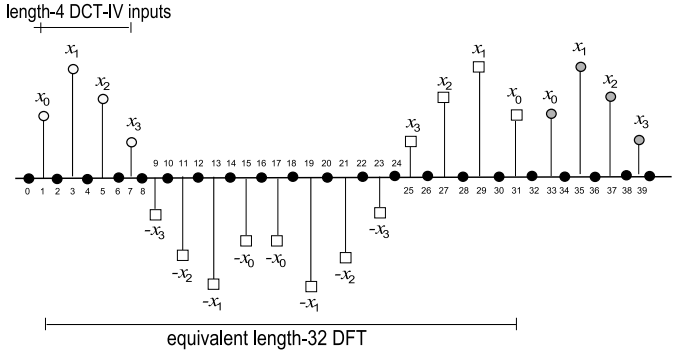}\par\end{centering}

\caption{\label{fig:DCT-IV-from-DFT}A DCT-IV of length $N=4$ (open dots
$x_{0},x_{1},x_{2},x_{3}$) is equivalent to a size $8N=32$ DFT via
interleaving with zeros (black dots) and extending to an odd-even-odd
(square dots) periodic (gray dots) sequence.}
\end{figure}

This is illustrated by an example, for $N=4$, in Fig\@.~\ref{fig:DCT-IV-from-DFT}.
The original four inputs of the DCT-IV are shown as open dots, which
are interleaved with zeros (black dots) and extended to an odd-even-odd
(square dots) periodic (gray dots) sequence of length $8N=32$ for
the corresponding DFT. Referring to eq.~(\ref{eq:DCT-IV from DFT}),
the output of the DCT-IV is given by the first $N$ odd-index outputs
of the corresponding DFT (with a scale factor of $1/4$). (The type-IV
DCT is distinguished from the other types by the fact that it is even
around the left boundary of the original data while it is odd around
the right boundary, and the symmetry points fall halfway in between
pairs of the original data points.) We will refer, below, to this
figure in order to illustrate what happens when an FFT algorithm is
applied to this real-symmetirc zero-interleaved data.%

\subsection{DCT-IV from DCT/DST-III\label{sub:DCT-IV-from-DCT-III}}

For a DCT-IV of size $N$, our strategy is to directly apply the new
FFT algorithm to the equivalent DFT of size $\tilde{N}=8N$, and to
discard the redundant operations from each stage. As it turns out,
the sub-transforms after one step of this algorithm are actually scaled-output
type-III DCTs and DSTs. This is closely related to a well-known algorithm
to express a DCT-IV in terms of a half-size DCT-III and DST-III \cite{Wang85}.
In this section, we derive this reduction to a DCT-III for the new
FFT algorithm, and then in Sec.~\ref{sec:New-DCT-III} we derive
a new algorithm for the scaled-output DCT-III. The scaled output DST-III
algorithm can be easily re-expressed in terms of the DCT-III, as is
presented in Sec\@.~\ref{sub:DST-III-from-DCT-III}. Here, we define
the DCT-III and DST-III, respectively, by the (unnormalized) equations:\begin{equation}
C_{k}^{\mathrm{III}}=\sum_{n=0}^{N-1}x_{n}\cos\left[\frac{\pi}{N}n\left(k+\frac{1}{2}\right)\right],\label{eq:DCT-III}\end{equation}
\begin{equation}
S_{k}^{\mathrm{III}}=\sum_{n=1}^{N}x_{n}\sin\left[\frac{\pi}{N}n\left(k+\frac{1}{2}\right)\right].\label{eq:DST-III}\end{equation}

Starting with the DFT of length $\tilde{N}$ in eq\@.~(\ref{eq:DFT}),
the new split-radix FFT algorithm splits it into three smaller DFTs:
$U_{k}=\dft(\tilde{x}_{2n_{2}})$ of size $\tilde{N}/2$, as well
as the scaled transforms $Z_{k}=\frac{1}{s_{\tilde{N}/4,k}}\dft(2\tilde{x}_{4n_{4}+1})$
and $Z_{k}'=\frac{1}{s_{\tilde{N}/4,k}}\dft(2\tilde{x}_{4n_{4}-1})$
of size $\tilde{N}/4$ (including a factor of 2 for convenience).
These are combined via:

\begin{equation}
X_{k}=U_{k}+\omega_{\tilde{N}}^{k}s_{\tilde{N}/4,k}Z_{k}/2+\omega_{\tilde{N}}^{-k}s_{\tilde{N}/4,k}Z_{k}'/2.\label{eq:new-conjugate-pair}\end{equation}
Here, where $\tilde{x}_{n}$ comes from the DCT-IV as in Sec\@.~\ref{sub:DCT-IV-from-DFT},
the even-indexed elements in $\tilde{x}_{n}$ are all zero, so $U_{k}=0$.
Furthermore, by the even symmetry of $\tilde{x}_{n}$, we have $Z_{k}'=Z_{k}^{*}$
(complex conjugate of $Z_{k}$). Thus, we have $C_{k}^{\mathrm{IV}}=X_{2k+1}=\Re(\omega_{8N}^{2k+1}s_{2N,2k+1}Z_{2k+1})$.
We will now show that $Z_{k}$ is given by combining a DCT-III and
a DST-III.

\begin{figure}
\begin{centering}\includegraphics[width=1.0\columnwidth]{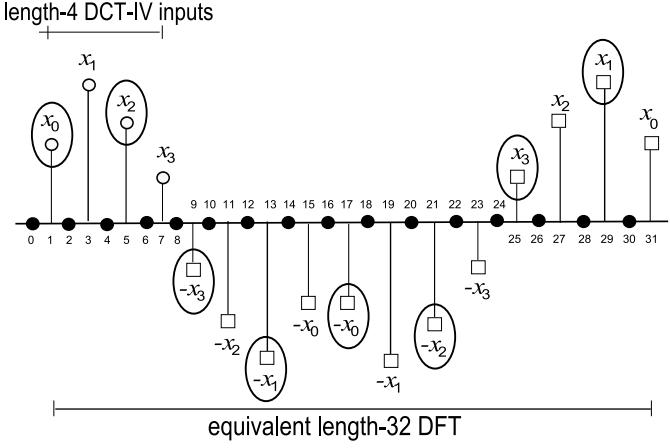}\par\end{centering}

\caption{\label{fig:fast-dct-IV}The DCT-IV of size 4 (open dots) is computed,
in a split-radix conjugate-pair FFT of the $\tilde{N}=32$ extended
data $\tilde{x}_{n}$ from Fig.~\ref{fig:DCT-IV-from-DFT}, via the
DFT $Z_{k}$ of the circled points $\tilde{x}_{4n+1}$, which is an
anti-periodic sequence of length $2N=8$.}
\end{figure}

In order to calculate $Z_{k}$, we denote for simplicity the inputs
of this subtransform by $z_{k}=2\tilde{x}_{4k+1}$ for $0\leq k<2N$.
Since $Z_{k}$ is the output of a real-input DFT of size $2N$, we
have $Z_{2N-k}=Z_{k}^{*}$. Thus, for any $0\leq k<N/2$, $Z_{2(N-1-k)+1}=Z_{2N-(2k+1)}=Z_{2k+1}^{*}$.
However, there is an additional redundancy in this transform that
we must exploit: by inspection of the construction of $\tilde{x}_{n}$
and by reference to Fig\@.~\ref{fig:fast-dct-IV}, we see that the
inputs $z_{k}$ are actually a real \emph{anti-}periodic sequence
of length $N$ (which becomes periodic when it is extended to length
$2N$). We must exploit this symmetry in order to avoid wasting operations.%

In particular, by using the anti-periodic symmetry of $z_{k}$, we
can write the DFT of length $2N$ as a single summation of length
$N$: \begin{eqnarray*}
Z_{2k+1} & = & \frac{1}{s_{2N,2k+1}}\sum_{n=0}^{2N-1}\omega_{2N}^{n(2k+1)}z_{n}\\
 & = & \frac{1}{s_{2N,2k+1}}\left(\sum_{n=0}^{N-1}\omega_{2N}^{n(2k+1)}z_{n}\right.\\
 &  & \quad\left.+\,\sum_{n=0}^{N-1}\omega_{2N}^{(n+N)(2k+1)}z_{n+N}\right)\\
 & = & \frac{2}{s_{2N,2k+1}}\sum_{n=0}^{N-1}\omega_{2N}^{n(2k+1)}z_{n},\end{eqnarray*}
using the facts that $\omega_{2N}^{N(2k+1)}=\omega_{2}^{(2k+1)}=-1$
and that $z_{n+N}=-z_{n}$. Then, if we take the real and imaginary
parts of the third line above, we obtain precisely a DCT-III and a
DST-III, respectively, with outputs scaled by $1/s_{2N,2k+1}$. However,
these sub-transforms are actually of size $N/2$, because the symmetry
$\omega_{2N}^{(N-n)(2k+1)}=-\omega_{2N}^{-n(2k+1)}$ means that the
$z_{n}$ and $z_{N-n}$ terms merely add or subtract in the input:\begin{multline}
\Re Z_{2k+1}=\frac{2}{s_{2N,2k+1}}\sum_{n=0}^{N-1}\cos\left[\frac{\pi}{N/2}n\left(k+\frac{1}{2}\right)\right]z_{n}\\
=\frac{2z_{0}+\sum_{n=1}^{N/2-1}\cos\left[\frac{\pi n\left(k+\frac{1}{2}\right)}{N/2}\right]\cdot2(z_{n}-z_{N-n})}{s_{2N,2k+1}}\label{eq:real-Z}\end{multline}
\begin{multline}
\Im Z_{2k+1}=-\frac{2}{s_{2N,2k+1}}\sum_{n=0}^{N-1}\sin\left[\frac{\pi}{N/2}n\left(k+\frac{1}{2}\right)\right]z_{n}\\
=\frac{(-1)^{k}{2z}_{N/2}+\sum_{n=1}^{N/2-1}\sin\left[\frac{\pi n\left(k+\frac{1}{2}\right)}{N/2}\right]\cdot2(z_{n}+z_{N-n})}{-s_{2N,2k+1}}\label{eq:im-Z}\end{multline}
for any $0\leq k<N/2$. We can define two new sequences $w_{n}$ ($0\leq n<N/2$)
and $v_{n}$ ($1\leq n\leq N/2$) of length $N/2$ to be the inputs
of this DCT-III and DST-III, respectively:

\begin{equation}
w_{0}=2z_{0},\; w_{n>0}=2(z_{n}-z_{N-n})\label{eq:w_n}\end{equation}

\begin{equation}
v_{n<N/2}=-2(z_{n}+z_{N-n}),\; v_{N/2}={-2z}_{N/2}.\label{eq:v_n}\end{equation}
With this definition of $w_{n}$ and $v_{n}$, we can conclude from
eqs.~(\ref{eq:real-Z}--\ref{eq:im-Z}), and the definition of DCT-III
and DST-III, that the real part of $Z_{2k+1}$ is exactly a scaled-output
DCT-III of $w_{n}$, while the imaginary part of $Z_{2k+1}$ is exactly
a scaled-output DST-III of $v_{n}$. (The scale factors of $\pm2$
will disappear in the end: they combine with the 2 in $z_{n}=2\tilde{x}_{4n+1}$
to cancel the $1/4$ in the definition of $\tilde{x}_{n}$.)

Thus, we have shown that the first half of the sequence $Z_{2k+1}(0\leq k<N)$
can be found from a scaled-output DCT-III and DST-III of length $N/2$.
The second half of the sequence $Z_{2k+1}$ can be derived by the
relation $Z_{2(N-1-k)+1}=Z_{2k+1}^{*}$ obtained earlier. Given $Z_{2k+1}$,
the output of the original DCT-IV, $C_{k}^{\mathrm{IV}}$, can be
obtained by the formula $C_{k}^{\mathrm{IV}}=\Re(\omega_{8N}^{2k+1}s_{2N,2k+1}Z_{2k+1})$.
This algorithm, in which the computation of $z_{k}$ has been folded
into the computation of $w_{k}$ and $v_{k}$, is presented in Algorithm~\ref{alg:DCT-IV}.%
\begin{algorithm}
\begin{algorithmic}

\item[\textbf{function}] $C_{k=0..N-1}^{\mathrm{IV}} \leftarrow \nsplitdctIV{N}{x_n}$:

\COMMENT{computes DCT-IV}

\STATE $w_0 \leftarrow x_0$

\STATE $v_{N/2} \leftarrow x_{N-1}$

\FOR{$k=1$ to $N/2-1$}

\STATE $w_k \leftarrow x_{2k}+x_{2k-1}$

\STATE $v_k \leftarrow x_{2k-1}-x_{2k}$

\ENDFOR

\STATE $W_{k=0 \ldots N/2-1} \leftarrow \nsplitdctIII{N/2}{1}{w_n}$

\STATE $V_{k=0 \ldots N/2-1} \leftarrow \nsplitdstIII{N/2}{1}{v_n}$

\FOR{$k=0$ to $N/2-1$}

\STATE $Z_{2k+1} \leftarrow W_k + iV_k$

\STATE $Z_{2(N-1-k)+1} \leftarrow W_k - iV_k$

\ENDFOR

\FOR{$k=0$ to $N-1$}

\STATE $C_k^{\mathrm{IV}} \leftarrow \Re\left(\omega_{8N}^{2k+1}s_{2N,2k+1}Z_{2k+1}\right)$

\ENDFOR

\end{algorithmic}

\caption{\label{alg:DCT-IV}Fast DCT-IV algorithm in terms of scaled-output
DCT-III and DST-III, derived from Algorithm~\ref{alg:new-fft} by
discarding redundant operations.}
\end{algorithm}

In Algorithm~\ref{alg:DCT-IV}, $\nsplitdctIII N\ell u$ calculates
the DCT-III of $\{ w_{n}\}$ scaled by a factor of $1/s_{4\ell N,2k+1}$
for $\ell=0,1,2,4$, and will be presented in Sec.~\ref{sec:New-DCT-III}.
Similarly, $\nsplitdstIII N\ell v$ calculates the DST-III of $\{ v_{n}\}$
scaled by a factor of $1/s_{4\ell N,2k+1}$ for $\ell=0,1,2,4$, and
will be presented in Sec.~\ref{sub:DST-III-from-DCT-III} in terms
of $\nsplitdctIII N\ell u$.

If the scale factors $s_{2N,2k+1}$ are removed (set to 1) in Algorithm~\ref{alg:DCT-IV},
we recover a decomposition of the DCT-IV in terms of an ordinary (unscaled)
DCT-III and DST-III that was first described by Wang~\cite{Wang85}.
This well-known algorithm yields a flop count exactly the same as
previous results: $2N\log_{2}N+N$. (Wang obtained a slightly larger
count, apparently due to an error in adding his DCT-III and DST-III
counts.) The introduction of the scaling factors in Algorithm~\ref{alg:DCT-IV}
reduces the flop count by simplifying some of the multiplications
in the scaled DCT-III/DST-III compared to their unscaled versions,
as will be derived in Sec\@.~\ref{sec:New-DCT-III}. Note that,
in Algorithm~\ref{alg:DCT-IV}, multiplying by $\omega_{8N}^{2k+1}s_{2N,2k+1}$
does not require any more operations than multiplying by $\omega_{8N}^{2k+1}$,
because the constant product $\omega_{8N}^{2k+1}s_{2N,2k+1}$ can
be precomputed. Let $M_{S}(N)$ denote the number of flops saved in
$\nsplitdctIII N1u$ compared to the best-known unscaled DCT-III.
We shall prove in Sec.~\ref{sub:DST-III-from-DCT-III} that the same
number of operations, $M_{S}(N)$, can be saved in $\nsplitdstIII N1u$.
Thus, the total number of flops required by Alg\@.~\ref{alg:DCT-IV}
will be $2N\log_{2}N+N-2M_{S}(N/2)$. The formula for $M_{S}(N)$
will be derived in Sec.~\ref{sec:New-DCT-III}, leading to the final
DCT-IV flop count formula given by eq\@.~(\ref{eq:flop-count-dct-IV}).

\section{DCT-III from new FFT\label{sec:New-DCT-III}}

The type-III DCT (for a convenient choice of normalization) is defined
by eq.~(\ref{eq:DCT-III}) for $N$ inputs $x_{n}$ and $N$ outputs
$C_{k}^{\mathrm{III}}$. We will now follow a process similar to that
in the previous section for the DCT-IV: we first express $C_{k}^{\mathrm{III}}$
in terms of a larger DFT of length $4N$, then apply the new FFT algorithm
of Sec.~\ref{sub:newfft}, and finally discard the redundant operations
to yield an efficient DCT-III algorithm. The resulting algorithm matches
the DCT-III flop count of our previous publication \cite{ShaoJo07-preprint},
which improved upon classic algorithms by about 6\% asymptotically.
Unlike our previous DCT-III algorithm, however, the algorithm presented
here also gives us an efficient \emph{scaled-output} DCT-III, which
saves $M_{S}(N)$ operations over the classic DCT-III algorithms.

\subsection{DCT-III in terms of DFT\label{sub:DCT-III-from-DFT}}

In order to derive $C_{k}^{\mathrm{III}}$ from the DFT formula, one
can use the identity $\cos\frac{\pi\ell}{N}=\frac{1}{4}\left(\omega_{4N}^{2\ell}-\omega_{4N}^{2N-2\ell}-\omega_{4N}^{2N+2\ell}+\omega_{4N}^{4N-2\ell}\right)$
to write:\begin{align}
C_{k}^{\mathrm{III}} & =\sum_{n=0}^{N-1}x_{n}\cos\left[\frac{\pi}{N}n\left(k+\frac{1}{2}\right)\right]\nonumber \\
 & =x_{0}+\sum_{n=1}^{N-1}\frac{x_{n}}{4}\left[\omega_{4N}^{n(2k+1)}-\omega_{4N}^{(2N-n)(2k+1)}\right.\nonumber \\
 & \qquad\qquad\left.-\,\omega_{4N}^{(2N+n)(2k+1)}+\omega_{4N}^{(4N-n)(2k+1)}\right]\nonumber \\
 & =\sum_{n=0}^{4N-1}\tilde{x}_{n}\omega_{4N}^{n(2k+1)}\label{eq:DCT-III from DFT}\end{align}
where $\tilde{x}_{n}$ is a real-even sequence of length $\tilde{N}=4N$,
defined as follows for $0<n<N$:

\begin{equation}
\tilde{x}_{n}=\tilde{x}_{4N-n}=\frac{1}{4}x_{n}\label{eq:tilde1-III}\end{equation}

\begin{equation}
\tilde{x}_{2N-n}=\tilde{x}_{2N+n}=-\frac{1}{4}x_{n},\label{eq:tilde2-III}\end{equation}
with $\tilde{x}_{0}=x_{0}/2$, $\tilde{x}_{N}=0$, $\tilde{x}_{2N}=-x_{0}/2$,
and $\tilde{x}_{3N}=0$. (Notice that the definitions of $\tilde{N}$
and $\tilde{x}_{n}$ here are different from those in Sec.~\ref{sub:DCT-IV-from-DFT}.)

\begin{figure}
\begin{centering}\includegraphics[width=1.0\columnwidth]{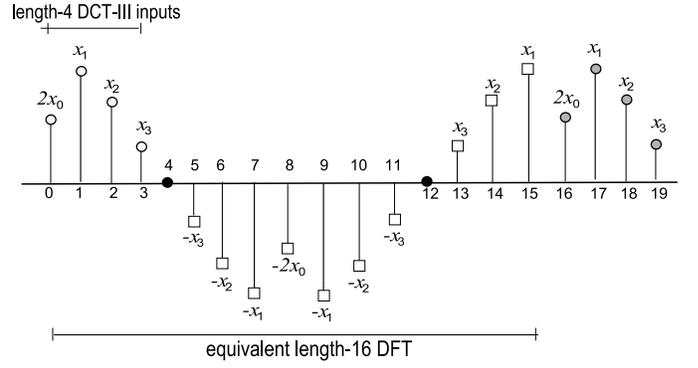}\par\end{centering}

\caption{\label{fig:DCT-III-from-DFT}A DCT-III of length $N=4$ (open dots
$x_{0},x_{1},x_{2},x_{3}$) is equivalent to a DFT of size $4N=16$
(scaled by a factor of $1/4$), via extending to an odd-even-odd (square
dots) periodic (gray dots) sequence and doubling the $x_{0}$ term.}
\end{figure}

This is illustrated by an example, for $N=4$, in Fig\@.~\ref{fig:DCT-III-from-DFT}.
This figure is very similar to Fig.~\ref{fig:DCT-IV-from-DFT}: both
of them are even around the points $n=0$ and $n=\tilde{N}/2$, and
are odd around the points $n=\tilde{N}/4$ and $n=3\tilde{N}/4$.
The difference from the DCT-IV is that these points of symmetry/anti-symmetry
are now data points of the original sequence, and so the data is not
interleaved with zeros as it was for the DCT-IV. We will refer, below,
to this figure in order to illustrate what happens when an FFT algorithm
is applied to this real-symmetirc data.%
.

\subsection{New (scaled) DCT-III algorithm}

In this subsection, we will apply the new FFT algorithm (Alg.~\ref{alg:new-fft})
to the corresponding DFT for a DCT-III of size $N$ as obtained in
Sec.~\ref{sub:DCT-III-from-DFT}. This process is similar to what
we did in Sec.~\ref{sub:DCT-IV-from-DCT-III}. We will see that a
DCT-III of size $N$ can be calculated by three subtransforms: a DCT-III
of size $N/2$, a DCT-III of size $N/4$, and a DST-III of size $N/4$.
The resulting algorithm for the DCT-III will have the same recursive
structure as in Alg.~\ref{alg:new-fft}: four mutually recursive
subroutines that compute the DCT-III with output scaled by different
factors. For use in the DCT-IV algorithm from Sec.~\ref{sub:DCT-IV-from-DCT-III},
we will actually use only three of these subroutines, because we will
only need a scaled-output DCT-III and not the original DCT-III.

When the new FFT algorithm is applied to the sequence $\tilde{x}_{n}$
of length $\tilde{N}=4N$ defined by eqs.~(\ref{eq:tilde1-III}--\ref{eq:tilde2-III}),
we get three subtransforms: of the sequences $\tilde{x}_{2n_{2}}$,
$\tilde{x}_{4n_{4}+1}$, and $\tilde{x}_{4n_{4}-1}$. The DFT of the
sequence $\tilde{x}_{2n_{2}}$ is equivalent to a size-$N/2$ DCT-III
of the original even-indexed data $x_{2n}$ , as can be seen from
Fig.~\ref{fig:DCT-III-from-DFT}. The subtransforms of $\tilde{x}_{4n_{4}+1}$
and $\tilde{x}_{4n_{4}-1}$ have exactly the same properties as the
corresponding subtransforms of the DCT-IV as described in Sec.~\ref{sub:DCT-IV-from-DCT-III}
(except that the length of the subtransform $\tilde{x}_{4n_{4}+1}$
here is $N$ instead of $2N$ as in the DCT-IV case). That is, we
denote the DFT of $2\tilde{x}_{4n_{4}+1}$ by $Z_{k}$, and this combines
with the DFT $Z_{k}^{'}=Z_{k}^{*}$ of $2\tilde{x}_{4n_{4}-1}$ to
yield a $\Re(\omega_{4N}^{2k+1}Z_{2k+1})$ term in the output as before.
And, as before, the inputs of $\tilde{x}_{4n_{4}+1}$ are anti-periodic
with length $N/2$. In consequence, we can apply the result derived
in Sec.~\ref{sub:DCT-IV-from-DCT-III} to conclude that these two
subtransforms can be found from a DCT-III of size $N/4$ and a DST-III
of size $N/4$. The corresponding inputs of the DCT-III and DST-III,
$w_{n}$ ($0\leq n<N/4$) and $v_{n}$ ($1\leq n\leq N/4$), are defined
as follows {[}compare to eqs.~(\ref{eq:w_n}--\ref{eq:v_n})]:

\begin{equation}
w_{0}=2z_{0},\; w_{n>0}=2(z_{n}-z_{N/2-n})\label{eq:w_n_DCT-III}\end{equation}

\begin{equation}
v_{n<N/4}=-2(z_{n}+z_{N/2-n}),\; v_{N/4}=-2z_{N/4},\label{eq:v_n-DCT-III}\end{equation}
where $z_{n}=2\tilde{x}_{4n+1}$ for $0\leq n<N/4$. (Again, the factors
of 2 will cancel the factor of $1/4$ in the definition of $\tilde{x}_{4n+1}$.)
Therefore, the real part of $Z_{2k+1}$ is the DCT-III of $w_{n}$,
while the imaginary part is the DST-III of $v_{n}$. In summary, a
DCT-III of size $N$ can be calculated by a DCT-III of size $N/2$,
a DCT-III of size $N/4$, and a DST-III of size $N/4$. Without the
scaling factors $s$, this is equivalent to a decomposition derived
by Wang of a DCT-III of size $N$ into a DCT-III and a DCT-IV of size
$N/2$ \cite{Wang84}, in which the DCT-IV is then decomposed into
a DCT-III and a DST-III of size $N/4$ \cite{Wang85}.

The above discussion is independent of the scaling factor applied
to the output of the transform. So, for the various scale factors
in the different subroutines of Alg.~\ref{alg:new-fft}, we simply
scale the DCT-III and DST-III subtransforms in the same way to obtain
similar savings in the multiplications (as quantified in the next
section). This results in the new DCT-III algorithm presented in Algorithm~\ref{alg:DCT-III}.
(The base cases, for $N=1$ and $N=2$, are omitted for simplicity.)
\begin{algorithm}
\begin{algorithmic}

\item[\textbf{function}] $C_{k=0..N-1}^{\mathrm{III}} \leftarrow \nsplitdctIII{N}{\ell}{x_n}$:

\COMMENT{computes DCT-III / $s_{4\ell N,2k+1}$ for $\ell = 0,1,2$}

\STATE $w_0 \leftarrow x_1$

\STATE $v_{N/4} \leftarrow x_{N-1}$

\FOR{$k=1$ to $N/4-1$}

\STATE $w_k \leftarrow x_{4k+1}+x_{4k-1}$

\STATE $v_k \leftarrow x_{4k-1} - x_{4k+1}$

\ENDFOR

\STATE $U_{k_2=0 \ldots N/2-1} \leftarrow \nsplitdctIII{N/2}{2\ell}{x_{2n_2}}$

\STATE $W_{k_4=0 \ldots N/4-1} \leftarrow \nsplitdctIII{N/4}{1}{w_{n_4}}$

\STATE $V_{k_4=0 \ldots N/4-1} \leftarrow \nsplitdstIII{N/4}{1}{v_{n_4}}$

\FOR{$k=0$ to $N/4-1$}

\STATE $Z_{2k+1} \leftarrow W_k + i V_k$

\STATE $Z_{N-2k-1} \leftarrow W_k - i V_k$

\ENDFOR

\FOR{$k=0$ to $N/2-1$}

\STATE $C_k^{\mathrm{III}} \leftarrow U_k + \Re \left( t_{4N,2k+1} Z_{2k+1} \right)  \frac{s_{4N,2k+1}}{s_{4 \ell N,2k+1})}$

\STATE $C_{N-k-1}^{\mathrm{III}} \leftarrow U_k - \Re \left( t_{4N,2k+1} Z_{2k+1} \right)  \frac{s_{4N,2k+1}}{s_{4 \ell N,2k+1}}$

\ENDFOR

\vskip 3pt

\item[\textbf{function}] $C_{k=0..N-1}^{\mathrm{III}} \leftarrow \nsplitdctIII{N}{4}{x_n}$:

\COMMENT{computes DCT-III / $s_{16N,2k+1}$}

\STATE $w_0 \leftarrow x_1$

\STATE $v_{N/4} \leftarrow x_{N-1}$

\FOR{$k=1$ to $N/4-1$}

\STATE $w_k \leftarrow x_{4k+1}+x_{4k-1}$

\STATE $v_k \leftarrow x_{4k-1} - x_{4k+1}$

\ENDFOR

\STATE $U_{k_2=0 \ldots N/2-1} \leftarrow \nsplitdctIII{N/2}{2}{x_{2n_2}}$

\STATE $W_{k_4=0 \ldots N/4-1} \leftarrow \nsplitdctIII{N/4}{1}{w_{n_4}}$

\STATE $V_{k_4=0 \ldots N/4-1} \leftarrow \nsplitdstIII{N/4}{1}{v_{n_4}}$

\FOR{$k=0$ to $N/4-1$}

\STATE $Z_{2k+1} \leftarrow W_k + i V_k$

\STATE $Z_{N-2k-1} \leftarrow W_k - i V_k$

\ENDFOR

\FOR{$k=0$ to $N/2-1$}

\STATE $C_k^{\mathrm{III}} \leftarrow \left[ U_k + \Re \left( t_{4N,2k+1} Z_{2k+1} \right) \right]  \frac{s_{4N,2k+1}}{s_{16N,2k+1}}$

\STATE $C_{N-k-1}^{\mathrm{III}} \leftarrow \left[ U_k - \Re \left( t_{4N,2k+1} Z_{2k+1} \right) \right]  \frac{s_{4N,2k+1}}{s_{16N,2N+2k+1}}$

\ENDFOR

\end{algorithmic}

\caption{\label{alg:DCT-III}New DCT-III algorithm of length $N$. The sub-transforms
$\nsplitdctIII N\ell x$ for $\ell\neq0$ are scaled by $s_{4\ell N,2k+1}$,
respectively, while $\ell=0$ is the final unscaled DFT ($s_{0,2k+1}=1$).}
\end{algorithm}

Just as in Sec\@.~\ref{sub:newfft}, although $\ell=0,1,2$ are
presented here in a compact form by a single subroutine $\ \nsplitdctIII N\ell x$,
in practice they would have to be implemented as separate subroutines
in order to exploit the special cases of the multiplicative constants
$s_{4N,2k+1}/s_{4\ell N,2k+1}$, similar to our FFT \cite{Johnson07}.

\subsection{DST-III from DCT-III \label{sub:DST-III-from-DCT-III}}

The DST-III and DCT-III are closely related. In particular, a DST-III
can be obtained from a DCT-III, with the same number of flops, by
reversing the inputs and multiplying every other output by $-1$ \cite{Wang82,Chan90,Lee94,ShaoJo07-preprint}:\begin{align}
S_{k}^{\mathrm{III}} & =\sum_{n=1}^{N}x_{n}\sin\left[\frac{\pi}{N}\left(k+\frac{1}{2}\right)n\right]\nonumber \\
 & =(-1)^{k}\sum_{n=0}^{N-1}x_{N-n}\cos\left[\frac{\pi}{N}\left(k+\frac{1}{2}\right)n\right].\label{eq:dst-III-from-dct-III}\end{align}
Similarly, one can derive a DST-III in terms of a DCT-III algorithm
for any scaling factor. Here, we present the algorithm $\nsplitdstIII N\ell v$
in terms of $\nsplitdctIII N\ell u$ for $\ell=0,1,2,4$. As we can
see from Alg.~\ref{alg:DST-III}, the new DST-III algorithm (with
scaled output) has exactly the same operation count as the corresponding
DCT-III algorithm (unary negations are not counted in the flops, because
they can be absorbed by converting additions into subtractions or
vice versa in the preceding DCT computation). This proves our previous
assertion that the numbers of operations saved in $\nsplitdctIII N1u$
and $\nsplitdstIII N1v$, compared to the known unscaled algorithms,
are both the same number $M_{S}(N)$.%
\begin{algorithm}
\begin{algorithmic}

\item[\textbf{function}] $S_{k=0..N-1}^{\mathrm{III}} \leftarrow \nsplitdstIII{N}{\ell}{x_n}$:

\COMMENT{computes DST-III $/s_{4\ell N,2k+1}$}

\FOR{$k=0$ to $N-1$}

\STATE $w_k \leftarrow x_{N-k}$

\ENDFOR

\STATE $C_{k=0 \ldots N-1}^{\mathrm{III}} \leftarrow \nsplitdctIII{N}{\ell}{w_n}$

\FOR{$k=0$ to $N-1$}

\STATE $S_k^{\mathrm{III}} \leftarrow (-1)^k C_k^{\mathrm{III}}$

\ENDFOR

\end{algorithmic}

\caption{\label{alg:DST-III}scaled-output DST-III algorithm of size $N$,
based on the scaled-output DCT-III algorithm which is presented in
Sec.~\ref{sec:New-DCT-III}, with the same operation count.}
\end{algorithm}

\section{Flop counts for DCT-III/IV}

First, we will show that Alg.~\ref{alg:DCT-III} gives the best previous
flop count $T(N)=2N\log_{2}N-N+1$ for the DCT-III if the scaling
factor $s$ is set to 1. Inspection of Algorithm~\ref{alg:DCT-IV}
yields a flop count $4N-2+2T(N/2)$ for the DCT-IV, and substituting
$T(N)$ gives the previous best flop count of $2N\log_{2}N+N$ for
the DCT-IV. Then, we will analyze how many operations are \emph{saved}
when the scaling factors are included.

If $s=1$, we can see from Alg.~\ref{alg:DCT-III} that a DCT-III
of size $N$ is decomposed into a DCT-III of size $N/2$, a DCT-III
of size $N/4$ and a DST-III of size $N/4$, and all four of our recursive
subroutines are identical (they only differed by $s$ factors). In
addition, $2(N/4-1)$ flops are required to obtain the sequences $w_{k}$
and $v_{k}$, and $5N/2$ flops are required to obtain the output
of the DCT-III from the outputs of the subtransforms. Therefore, we
obtain the recurrence relation for $T(N)$:

\begin{equation}
T(N)=T(N/2)+2T(N/4)+3N-2.\label{eq:T(N) recurrence}\end{equation}
The initial conditions for $T(N)$ can be determined easily. If $N=1$,
$y_{0}=x_{0}$. Therefore, $T(1)=0$. If $N=2$, $y_{0}=x_{0}+x_{1}/\sqrt{2}$
and $y_{1}=x_{0}-x_{1}/\sqrt{2}$. Therefore, $T(2)=3$. Solving eq.~(\ref{eq:T(N) recurrence})
with these initial conditions, we immediately obtain the following
result:

\begin{equation}
T(N)=2N\log_{2}N-N+1.\label{eq:T(N)}\end{equation}
This flop count is the same as the previous best flop count for DCT-III
algorithms \cite{Lee84,Vetterli84,Wang85,Duhamel86,Suehiro86,Hou87,Chan90,Arguello90,Lee94,Kok97,Takala00,Puschel03,Puschel03:dct}
prior to our work \cite{Johnson07,ShaoJo07-preprint}.

Since our DCT-III algorithm without scaling factors (i.e. with $s=1$)
obtains the same number of flops as the best previous DCT-III algorithms,
it only remains to determine how many operations are \emph{saved}
by including the scale factors. We now analyze this count of saved
flops by solving the appropriate recurrence relations. \emph{}In particular,
let $M(N)$, $M_{S}(N)$, $M_{S2}(N)$ and $M_{S4}(N)$ (where $N$
is a power of $2$) denote the number of operation saved (or spent,
if negative) in $\nsplitdctIII N\ell x$ for $\ell=0,1,2,4$, respectively,
compared to the corresponding unscaled DCT-III algorithm.

First, let us derive the recurrence relations for $M(N)$ and so on,
similar to the analysis of Alg.~\ref{alg:new-fft} \cite{Johnson07}.
The number of flops saved in $\nsplitdctIII N\ell x$ is the sum of
the flops saved in the subtransforms and the number of flops saved
in the loop to calculate the final results $C_{k}^{\mathrm{III}}$.
In $\nsplitdctIII N0x$, $5\cdot\frac{N}{2}$ flops are required in
the loop, as in the old unscaled algorithm. In $\nsplitdctIII N1x$,
only $4\cdot\frac{N}{2}$ flops are needed since either the real part
or the imaginary part of $t_{4N,2k+1}$is $1$. Thus, $N/2$ flops
in the loop are saved for $\ell=1$. In $\nsplitdctIII N2x$, $5\cdot\frac{N}{2}$
flops are again required in the loop. (In contrast, for Alg.~\ref{alg:new-fft}
the $\ell=2$ case required two more multiplications than the $\ell=0$
case because of the $k=0$ term \cite{Johnson07}, which is not present
here because $2k+1\neq0$.) In $\nsplitdctIII N4x$, $6\cdot\frac{N}{2}$
flops are required in the loop since $s_{16N,2k+1}\neq s_{16N,2k+1+2N}$
and hence we must multiply the two scale factors separately. This
means that we \emph{spend} $N/2$ extra multiplications in the $\ell=4$
case, which is counted as a negative term in $M_{S4}$. Thus, we have
the following relations:\begin{align}
M(N) & =M(N/2)+2M_{S}(N/4)\nonumber \\
M_{S}(N) & =M_{S2}(N/2)+2M_{S}(N/4)+N/2\nonumber \\
M_{S2}(N) & =M_{S4}(N/2)+2M_{S}(N/4)\nonumber \\
M_{S4}(N) & =M_{S2}(N/2)+2M_{S}(N/4)-N/2.\label{eq:op-count-gen}\end{align}
We next determine the number of flops saved (if any) for the base
cases, $N=1$ and $N=2$. When $N=1$, the unscaled algorithm computes
the output $y_{0}=x_{0}$, while the algorithms $\nsplitdctIII N\ell x$
calculate $y_{0}=(1/s_{4\ell,1})x_{0}$ which requires the same number
of flops for $\ell<2$ and one more multiplication for $\ell\geq2$:\begin{gather}
M(1)=M_{S}(1)=0,\nonumber \\
M_{S1}(1)=M_{S4}(1)=-1.\label{eq:op-count-1}\end{gather}
When $N=2$, we obtain scale factors $s_{4\ell N,2k+1}=s_{8\ell,2k+1}$.
For $\ell<4$, $s_{8\ell,1}=s_{8\ell,3}$, while $s_{8\ell,1}\neq s_{8\ell,3}$
for $\ell=4$. The unscaled algorithm calculates the output $y_{0,1}=x_{0}\pm\sqrt{1/2}x_{1}$,
where $3$ flops are required. For $\ell=0,1,2$, the algorithms $\nsplitdctIII N\ell x$
calculates $y_{0,1}=(x_{0}\pm\sqrt{1/2}x_{1})/s_{8\ell,1}$, which
requires 3 flops for $\ell=0$ (where $s=1$), 3 flops for $\ell=1$
(where $s=1/\sqrt{2}$ and cancels one of the constants), and 4 flops
for $\ell=2$. For $\ell=4$, $\nsplitdctIII N4x$ calculates $y_{0}=(x_{0}+\frac{1}{\sqrt{2}}x_{1})/s_{32,1}$
and $y_{1}=(x_{0}-\frac{1}{\sqrt{2}}x_{1})/s_{32,3}$, where $5$
flops are required. Thus, we have\begin{gather}
M(2)=M_{S}(2)=0,\nonumber \\
M_{S2}(2)=-1,\nonumber \\
M_{S4}(2)=-2.\label{eq:op-counts-2}\end{gather}
With these base cases, one can solve the recurrences (\ref{eq:op-count-gen})
by standard generating-function methods \cite{TAOCP-I} to obtain:\begin{equation}
M_{S}(N)=\frac{1}{9}N\log_{2}N-\frac{1}{27}N+\frac{1}{9}(-1)^{\log_{2}N}\log_{2}N+\frac{1}{27}(-1)^{\log_{2}N}.\label{eq:MS(N)}\end{equation}
Recall from Sec.~\ref{sub:DCT-IV-from-DCT-III} that $2M_{S}(N/2)$
flops can be saved in the new DCT-IV algorithm compared to the best
previous algorithms, resulting in a total flop count of $2N\log_{2}N+N-2M_{S}(N/2)$.
This gives the DCT-IV flop count in eq\@.~(\ref{eq:flop-count-dct-IV})
for Algorithm~\ref{alg:DCT-IV}. This expression for the flop count
of the new DCT-IV algorithm matches the results that were derived
by automatic code generation for small $N$ \cite{Johnson07}, as
expected.

In general, as was discussed in our other work on the DCT-II/III
\cite{ShaoJo07-preprint}, the number of multiplications may change
depending upon the normalization chosen. For the DCT-IV, a common
normalization choice is to multiply by $\sqrt{2/N}$, which makes
the transform unitary, but this does not change the number of flops
because the normalization can be absorbed into the $\omega_{8N}^{2k+1}s_{2N,2k+1}$
factor (which is $\neq1$ for all $k$). On the other hand, if one
is able to scale every output of the DCT-IV individually, for example
if the scale factor can be absorbed into a subsequent computational
step, then the best choice in the present algorithm seems to be to
scale by $1/s_{8N,2k+1}$. This choice of scale factor will transform
$\omega_{8N}^{2k+1}s_{2N,2k+1}$ into $t_{8N,2k+1}$ in Algorithm~\ref{alg:DCT-IV},
which can be multiplied in one fewer multiplication, saving $N$ multiplications
overall. Similarly, one can save $N$ multiplications for a scaled-\emph{input},
unscaled-output DCT-IV, since the scaled-output DCT-IV can be transformed
into a scaled-input DCT-IV by network transposition \cite{CrochiereOp75}
without changing the number of flops \cite{ShaoJo07-preprint}.

\section{DST-IV from DCT-IV}

The (unnormalized) DST-IV is defined as: \begin{equation}
S_{k}^{\mathrm{IV}}=\sum_{n=0}^{N-1}x_{n}\sin\left[\frac{\pi}{N}\left(n+\frac{1}{2}\right)\left(k+\frac{1}{2}\right)\right]\label{eq:dstIV}\end{equation}
for $k=0,\ldots,N-1$. Although we could derive fast algorithms for
$S_{k}^{\mathrm{IV}}$ directly by treating it as a DFT of length
$8N$ with odd symmetry, interleaved with zeros, and discarding redundant
operations similar to above, it turns out there is a simpler technique.
The DST-IV is \emph{exactly} equivalent to a DCT-IV in which the outputs
are reversed and every other input is multiplied by $-1$ (or vice
versa) \cite{Chan90}: \begin{equation}
S_{N-1-k}=2\sum_{n=0}^{N-1}(-1)^{n}x_{n}\cos\left[\frac{\pi}{N}\left(n+\frac{1}{2}\right)\left(k+\frac{1}{2}\right)\right]\label{eq:dstIV-dctIV}\end{equation}
for $k=0,\ldots,N-1$. It therefore follows that a DST-IV can be computed
with the same number of flops as a DCT-IV of the same size, assuming
that multiplications by $-1$ are free---the reason for this is that
sign flips can be absorbed at no cost by converting additions into
subtractions or vice versa in the subsequent algorithmic steps. Therefore,
our new flop count (\ref{eq:flop-count-dct-IV}) immediately applies
to the DST-IV.

\section{MDCT from DCT-IV}

In this section, we will present a new modified DCT (MDCT) algorithm
in terms of our new DCT-IV algorithm with an improved flop count compared
to the best previously published counts. The key fact is that the
best previous flop count for an MDCT of $2N=2^{m}$ inputs and $N$
outputs was obtained by reducing the problem to a DCT-IV plus $N$
extra additions \cite{Malvar90,Malvar91,Chan96,Liu99}. Therefore,
our improved DCT-IV algorithm immediately gives an improved MDCT.
Similarly for the inverse MDCT, except that in that case no extra
additions are required.

An MDCT of length {}``$N$'' has $2N$ inputs $x_{n}$ ($0\leq n<2N$)
and $N$ outputs $C_{k}^{\mathrm{M}}$ ($0\leq k<N$) defined by the
formula (not including normalization factors): 

\begin{equation}
C_{k}^{\mathrm{M}}=\sum_{n=0}^{2N-1}x_{n}\cos\left[\frac{\pi}{N}\left(n+\frac{1}{2}+\frac{N}{2}\right)\left(k+\frac{1}{2}\right)\right].\label{eq:MDCT}\end{equation}
This is {}``inverted'' by the inverse MDCT (IMDCT), which takes
$N$ inputs $C_{k}^{\mathrm{M}}$ and gives $2N$ outputs $y_{n}$,
defined by (again not including normalization):\begin{equation}
y_{n}=\sum_{k=0}^{N-1}C_{k}^{\mathrm{M}}\cos\left[\frac{\pi}{N}\left(n+\frac{1}{2}+\frac{N}{2}\right)\left(k+\frac{1}{2}\right)\right].\label{eq:IMDCT}\end{equation}
These transforms are designed to operate on consecutive 50\%-overlapping
blocks of data, and when the IMDCTs of subsequent blocks are added
in their overlapping halves the resulting {}``time-domain aliasing
cancellation'' (TDAC) yields the original data $x_{n}$ \cite{Princen86,Johnson87}.
The MDCT is widely used in audio compression, where the overlapping
reduces artifacts from the block boundaries \cite{Painter00}.

The MDCT and IMDCT can be trivially re-expressed in terms of a DCT-IV
of size $N$ \cite{Malvar90,Malvar91,Chan96,Liu99}. Let us define
\[
\Xi_{n}=\cos\left[\frac{\pi}{N}\left(n+\frac{1}{2}\right)\left(k+\frac{1}{2}\right)\right],\]
which has the symmetry $\Xi_{2N+n}=\Xi_{2N-1-n}=-\Xi_{n}$. In terms
of $\Xi_{n}$, the MDCT becomes\begin{align*}
C_{k}^{\mathrm{M}} & =\sum_{n=0}^{2N-1}\Xi_{\frac{N}{2}+n}x_{n}\\
 & =\sum_{n=0}^{\frac{N}{2}-1}\left(\Xi_{\frac{N}{2}+n}x_{n}+\Xi_{\frac{3N}{2}-1-n}x_{N-1-n}\right.\\
 & \qquad\qquad\left.+\,\Xi_{2N-1-n}x_{\frac{3N}{2}-1-n}+\Xi_{2N+n}x_{\frac{3N}{2}+n}\right)\\
 & =\sum_{n=0}^{\frac{N}{2}-1}\Xi_{\frac{N}{2}+n}\left(x_{n}-x_{N-1-n}\right)\\
 & \qquad-\,\sum_{n=0}^{\frac{N}{2}-1}\Xi_{n}\left(x_{\frac{3N}{2}-1-n}+x_{\frac{3N}{2}+n}\right)\\
 & =\sum_{n=N/2}^{N-1}\Xi_{n}\left(x_{n-N/2}-x_{\frac{3N}{2}-1-n}\right)\\
 & \qquad-\,\sum_{n=0}^{N/2-1}\Xi_{n}\left(x_{\frac{3N}{2}-1-n}+x_{\frac{3N}{2}+n}\right).\end{align*}
But the final summation is simply a DCT-IV of the sequence $\tilde{x}_{n}$
defined by $\tilde{x}_{n}=-(x_{\frac{3N}{2}-1-n}+x_{\frac{3N}{2}+n})$
for $0\leq n<\frac{N}{2}$ and $\tilde{x}_{n}=x_{n-N/2}-x_{\frac{3N}{2}-1-n}$
for $\frac{N}{2}\leq n<N$. Therefore, given any algorithm for a DCT-IV,
the MDCT can be computed with at most $N$ extra additions. (Here,
we are not counting multiplication by $-1$, because negations can
be absorbed by converting additions into subtractions and vice versa
in subsequent computational steps.) Since the previous best flop count
for the DCT-IV was $2N\log_{2}N+N$ flops, this led to a flop count
of $2N\log_{2}N+2N$ for the MDCT \cite{Malvar90,Malvar91,Chan96,Liu99}.
Instead, we can use our new algorithm for the DCT-IV to immediately
reduce this flop count for the MDCT to eq\@.~(\ref{eq:flop-count-dct-IV})$+N$:\begin{equation}
\frac{17}{9}N\log_{2}N+\frac{58}{27}N+\frac{2}{9}(-1)^{\log_{2}N}\log_{2}N-\frac{4}{27}(-1)^{\log_{2}N}.\label{eq:flop-count-mdct}\end{equation}

The IMDCT requires almost no manipulation: it is already in the form
of a DCT-IV, except that we are evaluating the DCT-IV beyond the {}``end''
of the inputs. Since a DCT-IV corresponds to anti-symmetric data as
in Fig\@.~\ref{fig:DCT-IV-from-DFT}, this just means that we compute
the DCT-IV and obtain the IMDCT by storing the outputs and their mirror
image (multiplied by $-1$), shifted by $N/2$. So, the flop count
for the IMDCT is exactly the same as the flop count for the DCT-IV,
not counting negations. Any overall negation of an output can be eliminated
by converting a preceding addition to a subtraction (or changing the
sign of a preceding constant), but some of the (redundant) IMDCT outputs
are needed with both signs, which seems to imply that an explicit
negation is required. The latter negations are easily eliminated in
practice, however: an IMDCT is always followed in practice by adding
overlapping IMDCT blocks to achieve TDAC, so the negations simply
mean that some of these additions are converted into subtractions.

\section{Concluding remarks}

We have derived new algorithms for the DCT-IV, DST-IV, MDCT, and IMDCT
that reduce the flops for a size $N=2^{m}$ from $2N\log_{2}N+O(N)$ to
$\frac{17}{9}N\log_{2}N+O(N)$, representing the first improvement in
their flop counts for many years and stemming from similar
developments for the DFT \cite{Johnson07,Lundy07}. We do not claim
that these flop counts are the best possible, although we are not
currently aware of any way to obtain further reductions in either the
leading coefficient or in our $O(N)$ terms. It is possible that
further gains could be made by extending our recursive rescaling
technique to greater generality, for example. However, we believe that
such investigations will be most easily carried out in the context of
the DFT, since FFT algorithms (in terms of complex exponentials) are
typically much easier to work with than fast DCT algorithms (in terms
of real trigonometry), and any improved FFT algorithm will immediately
lead to similar gains for DCTs (and vice versa: any improved DCT leads
to an improved FFT) \cite{DuhamelVe90}. Another open question is
whether these new algorithms will lead to practical gains in
performance on real computers. This is a complicated and somewhat
ill-defined question, because performance characteristics vary between
machines and depend strongly on many factors besides flop counts---any
simple algorithms like the ones presented here require extensive
restructuring to make them efficient on real computers, just as
classic split-radix does not perform well without modification
\cite{FFTW05}. On the other hand, for small fixed $N$ where
straight-line (unrolled) hard-coded kernels are often employed in
audio and image processing (where the block size is commonly fixed),
we have demonstrated that automatic code-generation techniques (given
only the new FFT) can produce efficient DCT-IV (and MDCT, etc.)
kernels attaining the new operation counts, and that the performance
is sometimes improved at least slightly \cite{Johnson07}.

\section*{Acknowledgements}

This work was supported in part by a grant from the MIT Undergraduate
Research Opportunities Program. The authors are also grateful to M.
Frigo, co-author of FFTW with SGJ \cite{FFTW05}, for many helpful
discussions.

\bibliographystyle{elsart-num}
\bibliography{dct}

\end{document}